# Influence of the Solar Luminosity on the Glaciations, Earthquakes and Sea Level Changes as Derived from Photography of Luminescence of Speleothems: Regional Sustainable Development Dimensions.

Yavor Y. Shopov
Section of Speleology and Faculty of Physics, University of Sofia, James Bouchier 5, Sofia 1164, Bulgaria
E-Mail:YYShopov@Phys.Uni-Sofia.BG

## Abstract

Glaciations were attributed to variations of the Earth's orbit (Milankovitch cycles). But the best ever dated paleoclimatic record (from a speleothem from Devils Hole, Nevada) demonstrated that the end of the last glacial period (termination II) happened 10 000 years before the one suggested by the orbital variations (Winograd et al, 1988, 1992), i.e. the result appeared before the reason. This fact suggests that there is something wrong in the theory. Glaciations and deglaciations drive changes of the sea level. They are extremely important for human life in coastal regions, because rise of the sea level with few meters will cause flooding of very large regions of land and will reduce significantly arable land of several countries. We need to know the detailed mechanisms of the deglaciations in order to develop proper sustainable management policy and actions to secure sustainable development of these regions. These actions are important not only for sustainable management of the caves (which provide the data for coming disasters), but for survival of large regions of land and its population.

Calcite speleothems luminescence of organics depends exponentially upon soil temperatures that are determined primarily by the solar radiation. So the microzonality of luminescence of speleothems may be used as an indirect Solar Insolation (radiation) proxy index. We obtained luminescence solar insolation proxy records in speleothems (from Jewel Cave, South Dakota, US and Duhlata cave, Bulgaria). These records exhibit very rapid increasing of the solar insolation at 139 kyrs BP responsible for the termination II (the end of the last glaciation) and demonstrate that solar luminosity variations contribute to Earth's heating almost as much as the orbital variations of the Earth's orbit (Milankovitch cycles). The most powerful cycle of the solar luminosity (11500 yrs) is responsible for almost 1/2 of the variations in solar insolation experimental records. Solar luminosity and orbital variations both cause variations of the solar insolation affecting the climate by the same mechanism.

Changes in the speed of Earth's rotation during glacial- interglacial transitions produce fracturing of the Earth's crust and major earthquakes along the fractures. The intensity of this process is as higher as faster is the change of the sea level and as higher is its amplitude. Much higher dimensions of this process may be caused by eruptive increasing of solar luminosity, which may be caused only by collision of large asteroids with the Sun (Shopov et al, 1996, 1997). Such collision may cause "Bible Deluge" type of event. Humanity even now is not prepared to face such catastrophic disaster, but cave research combined with astronomical observations can help to predict it and to suggest actions to reduce the damage cause by it.

## Reliability of the Orbital Theory

M.Milankovitch (1920) demonstrated that orbital variations of the Earth's orbit cause significant variations of the amount of solar radiation received by the Earth's surface (solar insolation-SI) and assumed the idea that glacial periods (ice ages) are result of such variations.

Recent measurements (Winograd et al, 1988, 1992) of a cave deposit from Devils Hole (DH), USA (which is the best dated paleoclimatic record) demonstrated that the end of the former glaciation (Termination II) came 10 kyrs before that suggested by the orbital theory. Some scientists consider this as a denial of the orbital origin of glaciations, because it demonstrates that the result appears far before the reason.

The Orbital theory has 2 presumptions:
1. That the solar luminosity is constant during geological periods of time.
2. That the Earth behaves as an absolutely solid body independently of the orbital variations.

Recent studies demonstrate that both these presumptions are not precise. Direct satellite measurements of the solar constant demonstrated that it varies with time as much as 0.4% during the observation time span (Hickey et al., 1980), but there are experimental data suggesting that it varied much greater during geological periods of time (Stuiver & Braziunas, 1989).

Formation and melting of the ice sheets affect the Sea Level, but they influence also the rotation speed of the Earth (Moerner, 1993) and the orbital variations (Bills, 1998). The orbital theory (Berger, Loutre 1992) does not consider these influences.

Increasing of the ice volume and related sea level change during glaciations produces changes in the inertial moment of the Earth and resulting changes in the speed of the Earth's rotation (Tenchov et al., 1993). These changes must affect in some degree the amplitude and may be even the period of the orbital variations. Orbital variations cause also some deformation of the solid Earth and redistribution of the Ocean masses (Moerner, 1976, 1983). In result theoretical curves can be used only for qualitative reference. For quantitative correlation it is necessary to use experimental records of the solar insolation, because they contain also variations of the solar luminosity and number of others not covered by the Orbital theory. So far such experimental records can be derived only from luminescence of speleothems, by photometry of photos of luminescence of polished sections of speleothems. So this way scientific studies of speleothems can provide unique information of major importance for understanding and qualitative assessment of the Global change and management of the sustainable development of coastal regions.

Such photos have not only great scientific value, but sometimes have also fantastic artistic and aesthetic value (photo 1), so they are well presented in the exhibition "A World of Fantasy Formed by Water and Time" of Cave Expo Korea 2002.

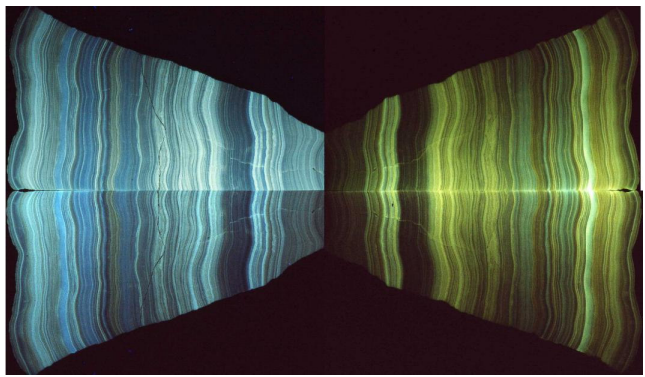

Photo 1. Luminescence of a 7 cm. long polished section of a flowstone from Jewel Cave, South Dakota, US. Fluorescence (left) and phosphorescence (right). Photo by Yavor Shopov, © Art@Net International Gallery.

The changes in the speed of Earth's rotation during glacial- interglacial transitions cause some deformation of the solid Earth (Tenchov et al., 1993) in order to adjust the shape of the Earth to the new rotation speed. This process cause fracturing of the Earth's crust and major earthquakes along the fractures. The intensity of this process is as higher as faster is the change of the sea level and as higher is its amplitude. Therefore century variations of the solar luminosity (Maunder and "Medieval Maximum" type of events) may cause earthquakes like glacial periods. Much higher dimensions of this process may be caused by eruptive increasing of solar luminosity, which may be caused only by collision of large asteroids with the Sun (Shopov et al, 1996, 1997). Such collision may cause "Bible Deluge" type of event.

## Experimental Methods

Calcite speleothems (stalagmites etc.) usually display luminescence (photo 1) which is produced by calcium salts of humic and fulvic acids derived from soils above the cave (White, Brennan, 1989). These acids are released by the decomposition of humic matter. Rates of decomposition depend exponentially upon soil surface temperatures that are determined primarily by solar infrared radiation (Shopov et al., 1997). So the microzonality of luminescence of speleothems can be used as an indirect Solar Activity (SA) index (Shopov et al, 1994). From a speleothem from Cold Water cave, (Iowa, USA) it was obtained extremely good correlation (fig.1) between variations of the past solar luminosity (SL) and the intensity of speleothem luminescence.

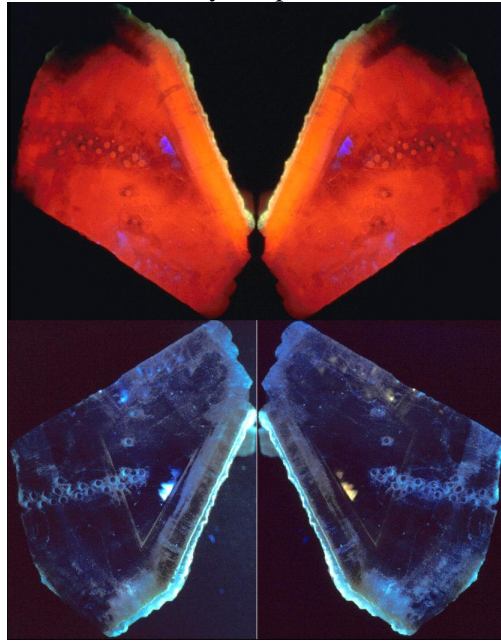

Photo 2. Luminescence of a section of a cave spar from a cave near Carlsbad cavern, US. Fluorescence under short-wave UV light (down left), under longwave UV light (down right) and phosphorescence (up). Mention blue phosphorescing gas inclusions of oil components, producing orange fluorescence if irradiated by longwave UV light. Photo by Yavor Shopov, ©Art@Net International Gallery.

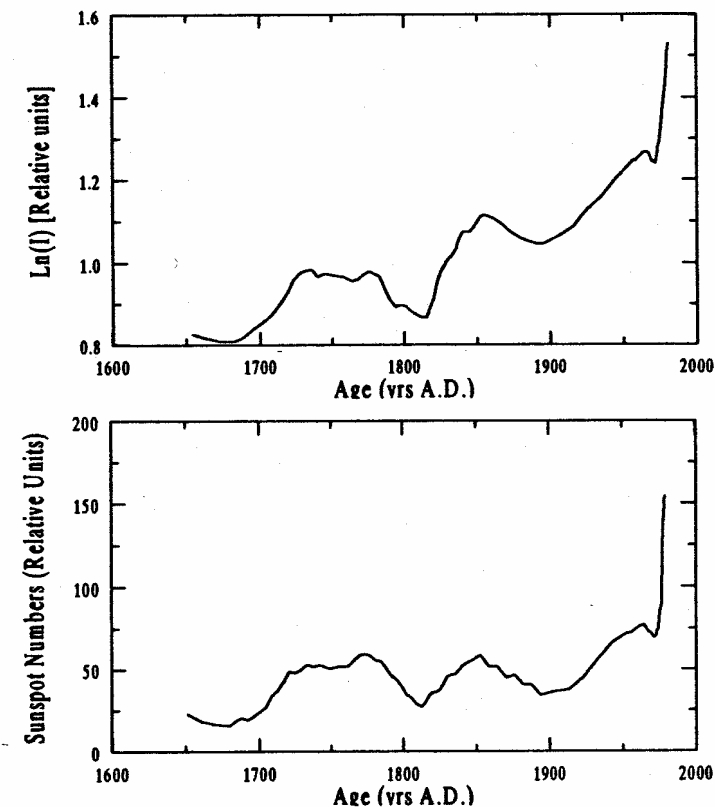

**Figure 1. (a)** 20 yr. average sunspot record (Solar Luminosity Sunspot index) since 1700 AD **(b)** Luminescent speleothem record from Cold Water cave, Iowa, USA (Shopov et al. 1996b).

Time series of the SA index "Microzonality of Luminescence of Speleothems" are obtained by Laser Luminescence Microzonal analysis (LLMZA) of cave flowstones. LLMZA allows measurement of luminescence time series with duration of hundreds of thousands years, and nevertheless the time step for short time series can be as small as 6 hours (Shopov et al., 1994) allowing resolution of 3 days.

### Luminescent Speleothem Proxy Record of Solar Insolation

We measured a luminescent solar insolation proxy record in a speleothem (JC11) from Jewel Cave, South Dakota, USA (Shopov et al.,1998, Stoykova et al., 1998). This record covers 89300-138600 yrs BP (fig.2b) with high resolution. It reveals determination of millennial and century cycles in the record.

This U/Th dated record exhibits a very rapid increasing in solar insolation at 139 kyrs ±5.5 kyrs BP responsible for the termination II. This increasing precedes the one suggested by the Orbital theory with about 10 kyrs and is due to the most powerful cycle of the solar luminosity with period of 11.5 kyrs superposed on the orbital variations curve (Stoykova et al., 1998). The Devils Hole paleoclimatic record (fig.2a) suggests that termination II had happened at 140 ±3 kyrs B.P. It follows precisely the shape of our experimental solar insolation record. This result is confirmed by an other U/Th dated luminescent solar insolation proxy record in a speleothem from a Duhlata cave, Bulgaria (fig.2c) 10 000 km far form the JC11 site. These records do not deny the orbital theory, but they suggest that the solar luminosity contribution to the solar insolation curves has been severely underestimated. The most powerful solar luminosity cycle in this record with period of 11.5 kyrs appears to be a bit more powerful than the precession cycle and a bit less than the total orbital component of the SI variations. Indeed theoretical curves (Berger, 1978, 1992) explain only about 1/2 of the signal in the existing proxy paleotemperature records (Imbrie et al.,1992, 1993).

### Solar Insolation, Solar Luminosity and Sea Level Variations

John Imbrie et al.(1985) demonstrated that orbital variations cause major changes of the global sea level, because of the melting of polar ice caps by the solar radiation. He even expressed units of the orbital variations in resulting sea level changes (in meters regarding the modern sea level). During the last glacial maximum 18000 years ago global sea level was 120 meters below the modern one. The reason for this is that water and ice adsorbs strongly infrared solar radiation, resulting in melting of the ice. Lower solar insolation during glaciations allows higher ratio between ice precipitation and melting resulting in increasing of ice accumulation and also in advance of the ice shields in direction to the equator. Melting of this ice during interglacials cause rising of the sea level.

Coastal caves and their speleothems serve as important indicators of the sea level changes.

### Speleothem Luminescence Proxy Records of Past Precipitation: Evidences for "The Deluge" in Speleothems.

We studied variations of the growth rate of a stalagmite from Duhlata cave, Sofia district, Bulgaria. Cave flowstones, stalactites and stalagmites, usually are built of calcium carbonate, obtained by partial dissolving of the bedrock over the cave. This solution precipitates a part of the dissolved carbonate in the form of cave speleothems. They grow continuously hundreds of thousand years. As bigger is precipitation, as higher is the amount of the bedrock dissolved and reprecipitated over the stalagmite and as faster is it's growth. Therefore speleothem growth rate is linearly proportional to the amount of precipitation in the past. A curve of the stalagmite growth rate variations, representing past precipitation has been obtained by dating of a sequence of points along the stalagmite growth axis. This way it is demonstrated, that around 7500 B.P. the stalagmite growth rate (averaged for 120 years) exceeds 50 times its recent value (Shopov et al, 1997b).

Shopov et al., (1996b,1997b) demonstrated, that the Bible Flood probably is with age equal to the beginning of the bible chronology (Creation of the World), i. e. 5500 years B.C., Presuming that the excess precipitation had fallen only within 1 year, this means a never seen rainfall (flood). Such event is described in the Bible, Greek mythology and the Scumerian epic Gilgamesh (compiled during III millennium B.C. on the base of more ancient legends). Such immemorial precipitation probably would lead to some temporary rising of the Black Sea level. Such rising at 5500 B.C. was recently suggested by an international team of scientists, lead by Dr.William Rayn and Walter Pittman from Columbia University, Palisades, New York. They demonstrated that the Black sea level rose with 150 meters in one year, flooding 160000 square kilometers. Before this event Black Sea was isolated fresh water lake. The Black Sea level rising itself cannot undoubtedly be related to the Flood, but combined with the never seen (during the human civilization) precipitation at that time definitely lead to the thought, that this phenomenon is namely the Bible Flood.

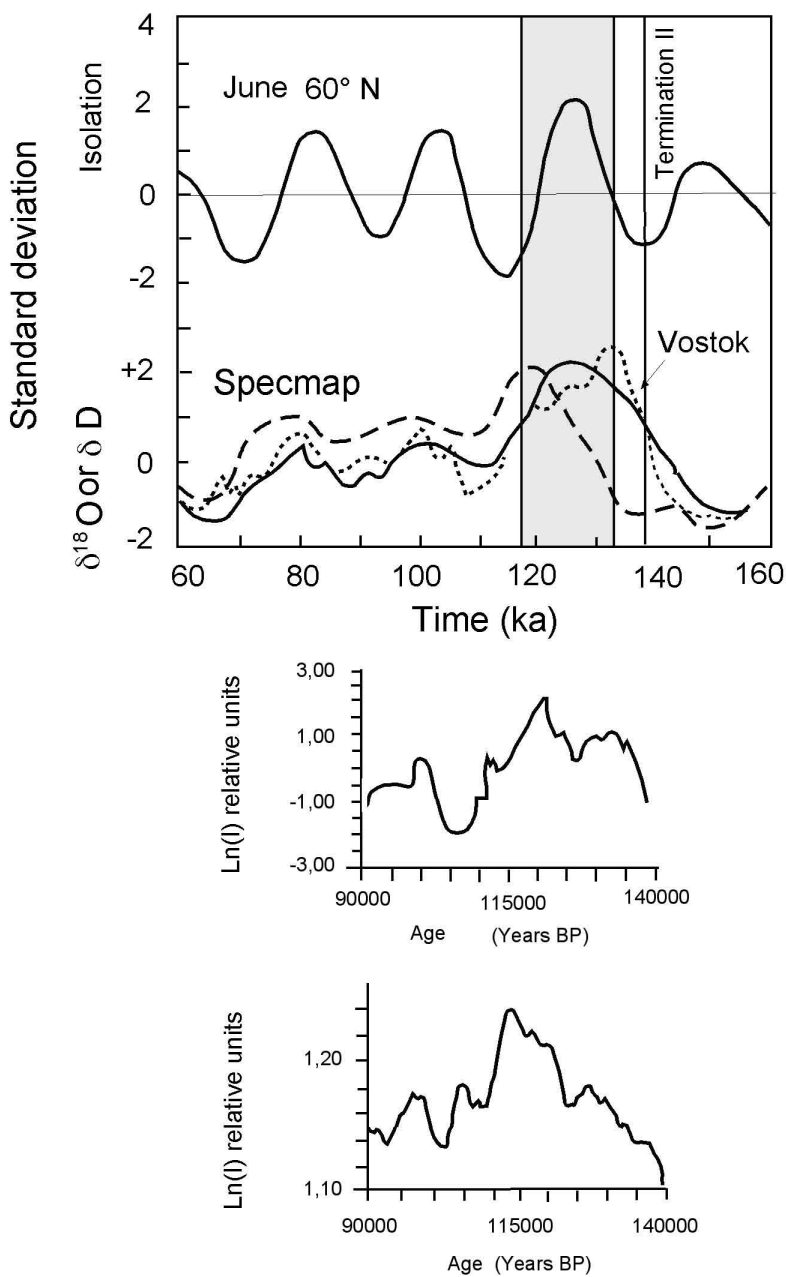

**Figure 2. (a)** The theoretical insolation curve compared to Devils Hole (DH-11), Vostok, and SPECMAP stack stable isotope curves (Winograd et al.,1992). Shading represents high sea level stands (at or above modern levels). **(b)** JC11 speleothem luminescence record from Jewel cave, US. **(c)** Duhlata Cave (Bulgaria), DC-2 luminescence Solar Insolation proxy record

Here we suggest a hypothesis for one possible mechanism of the Flood, consisting of the following: - It is known, that the Ocean level raised from 10000 years B.P. to the time of the Flood as a result of the glaciers melting. Black Sea had been isolated from the Ocean and its level had been much lower. In one moment the narrow band of land between the Mediterranean and Black Sea had broken down like a dam wall. This had resulted in flowing of giant masses of seawaters into the Black Sea basin. When it reached the opposite cost, then a giant wave had been formed (which probably was incomparably bigger that biggest tsunami known so far). This wave had destroyed everything on lands around the Black Sea even beyond the regions flooded by the sea level rise. The never seen precipitation at that time had contributed to the rising of the sea level (Moerner (1988) found a rapid rising of the sea level with 17 m between 8000 and 7500 B.P.) and maybe caused the final rising which had turned the Mediterranean Sea over the edge to flood the Black Sea region. Any known Earth force can not produce such precipitation. It requires rapid increasing in evaporation of water, but there are no evidences of rapid warming during the Flood. So the only possible reason for such evaporation is increasing of the solar luminosity with several %. Water absorbs strongly the infrared solar radiation, which cause melting of glaciers and evaporation of water. But evaporation cause cooling of the system (refrigerator effect). Solar luminosity usually remains rather steady (Solar constant). Such higher solar radiation can be produced only by explosion of a comet or asteroid in the solar atmosphere. Such explosion (like Tungussian meteorite) can cause a major mixing of parts of solar layers and appearance of warmer solar matter from the depth to the solar surface. Solar luminosity increases with forth degree of the temperature of the solar surface, so it should increase significantly immediately after the collision. Probably several years would be necessary for recovery of the Sun from such catastrophic event.

Such rapid melting of the ice sheets and rising of the sea level should cause unusually rapid change of the rotation speed of the Earth and would produce major earthquakes. Probably one of them broke the narrow band of land separating Black Sea from the Mediterranean Sea and caused the flooding of the Black Sea basin. Humanity even now is not prepared to face such catastrophic disaster, but cave research combined with astronomical observations can help to predict it and to suggest actions to reduce the damage cause by it.

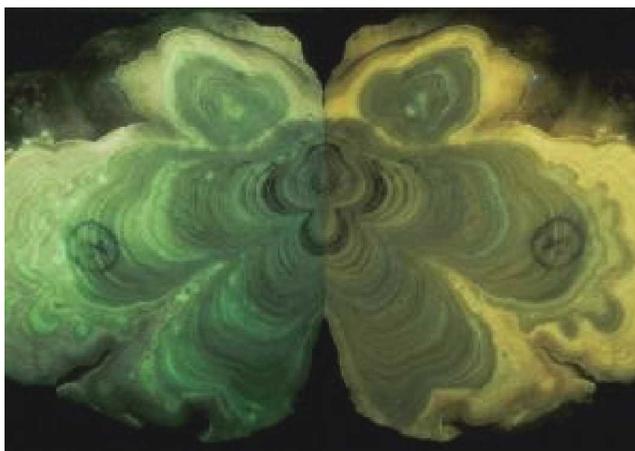

Photo 3. Luminescence of a section of a calcite coralloid from a cave near Irkutsk, Russia. Fine fluorescence banding under short-wave UV light (left) is produced by uranium impurities in the speleothem. Fine fluorescence banding under longwave UV light (right) is produced by rare earth elements in the same sample. Photo by Yavor Shopov.

**Pollution and migration of toxic compounds indicated by speleothem luminescence**
Usually researchers attribute all luminescence in calcite speleothems to organics without any reason to do so. Detailed spectral measurements of the luminescence are absolutely necessary to determine luminescent compounds in any speleothem. In many samples all or a significant part of the luminescence is produced by inorganic ions. Sometimes they even have annual banding (photo 3) due to variations of acidity of the karst waters, causing variations of the solubility of some pollutants or toxic compounds (Shopov, 1997). Uranium compounds have such migration behavior. On photo 3 is demonstrated fine fluorescence banding produced by uranium impurities in the speleothem under short-wave UV light. Fine fluorescence banding under longwave UV light is produced by rare earth elements in the same sample. Phosphorescence of this sample (not shown) suggests that there are no any luminescent organics in the middle (darker) part of the speleothem.

## Conclusions

Solar luminosity variations contribute to Earth's heating almost as much as the variations of the Earth's orbit (Milankovitch cycles). Their most prominent cycle (with period of 11500 yrs) must be also taken into account for a proper explanation of the timing of the deglaciations.

Speleothem records (being the best-dated paleoclimatic records) may serve as a reliable tool for studying the mechanisms of formation and precise timing of glacial periods.

Glaciations and deglaciations drive changes of the sea level. They are extremely important for human life in coastal regions, because rise of the sea level with few meters will cause flooding of very large regions of land and will reduce significantly arable land of several countries. We need to know the detailed mechanisms of the deglaciations in order to develop proper sustainable management policy and actions to secure sustainable development of these regions. These actions are important not only for sustainable management of the caves (which provide the data for coming disasters), but for survival of large regions of land and its population. Cave research is necessary to predict coming major global climatic disasters and to suggest proper policy and actions to reduce the damage caused by it.

Acknowledgements
I thank to Diana A. Stoykova, Ludmil Tsankov, Michael Sanambria and Leonid N. Georgiev for their priceless contribution to this research and especially to Derek Ford for his samples and their U/Th dating.
This research was funded by Bulgarian Science Foundation by research grant 811/98 to Y. Shopov